\documentstyle[12pt]{article}

 \newcommand{\be}{\begin{equation}}
 \newcommand{\ee}{\end{equation}}
 \newcommand{\ba}{\begin{eqnarray}}
 \newcommand{\ea}{\end{eqnarray}}
 
 \newcommand{\del}{\partial}

\def\la{\lambda}

\def\C{\tilde C}

\newcommand{\expo}{\exp \lef \{ - \int d^3z }

\newcommand{\lef}{\left}

\newcommand{\ri}{\right}

\newcommand{\cl}{{\cal L}}

\newcommand{\fr}{\frac}

\begin{document}

\begin{titlepage}

\topmargin -15mm

\rightline{\bf UFRJ-IF-FPC-007/96}

\vskip 10mm

\centerline{ \LARGE\bf Massive Quantum Vortex Excitations }
\vskip 2mm
\centerline{ \LARGE\bf in a Pure Gauge Abelian Theory in 2+1D}

    \vskip 1.0cm

    \centerline{\sc E.C.Marino }

     \vskip 0.6cm
     
\centerline{\it Instituto de F\'\i sica, Universidade Federal 
do Rio de Janeiro } 
    
\centerline{\it Cx.P. 68528, Rio de Janeiro, RJ 21945-970, Brasil} 
\vskip 0.4cm
\centerline{\sc and}
\vskip 0.4cm 
\centerline{\sc Fl\'avio I. Takakura}
\vskip 0.4cm
\centerline{\it Departamento de F\'\i sica, 
Universidade Federal de Juiz de Fora}
\centerline{\it Juiz de Fora, MG 36036-330, Brasil}
\centerline{\it and}
\centerline{\it Department of Physics and Astronomy, Pittsburgh University}
\centerline{\it Pittsburgh, PA 15260, USA}
\vskip 1.0cm

\begin{abstract} 
 
We introduce and study a pure gauge abelian theory in 2+1D in which
massive quantum vortex states do exist in the spectrum of excitations.
This theory can be mapped in a three dimensional gas of point particles
with a logarithmic interaction, in the grand-canonical ensemble. We claim
that this theory is the 2+1D analog of the Sine-Gordon, the massive vortices
being the counterparts of Sine-Gordon solitons. We show that a symmetry
breaking, order parameter, similar to the vacuum expectation value of a
Higgs field does exist.

\end{abstract}

\vskip 3cm
$^*$ Work supported in part by CNPq-Brazilian National Research Council.
     E-Mail addresses: marino@if.ufrj.br; takakura@fisica.ufjf.br

\end{titlepage}

\hoffset= -10mm

\leftmargin 23mm

\topmargin -8mm
\hsize 153mm
 
\baselineskip 7mm
\setcounter{page}{2}

\section{Introduction}

Massive vortex excitations usually occur in gauge theories
coupled to a symmetry breaking Higgs field in the broken phase \cite{no}.
In this work, we introduce a three dimensional pure gauge 
abelian theory which presents a mechanism for the generation of mass for
quantum vortex excitations that does not require the presence of 
a Higgs field. This 
is completely analogous to the one
through which the Sine-Gordon solitons acquire a mass in two 
space-time dimensions \cite{cm}: if we turn off the cossine interaction
in the Sine-Gordon theory, we are left with a free massless field, 
in which the soliton operator creates massless excitations \cite{cm}.
The quantum ``soliton'' system can then be associated to an electrostatic
system of point charges such that the ``soliton'' correlation functions
are exponentials of the electrostatic energy of these charges \cite{odd}.
When we turn on the cossine interaction, these external charges 
associated with the solitons are imbibed in a gas of point (dual) charges in
the grand-canonical ensemble \cite{sgcg} and the effective 
interaction of the former charges is thereby modified, leading to a 
behavior of the correlation functions which indicates the solitons
become massive. 

We start by considering the vortex operator \cite{pol,vor,nv}
in a modified Maxwell theory  in 2+1D (described by (\ref{mnl})) 
which is the three dimensional analog of the free massless scalar 
field, in the sense that its quantum vortex states are massless.
It is rather suggestive that this theory has appeared for the first 
time in the bosonization of the free massless Dirac fermion field in 2+1D
\cite{bos} in analogy to its two dimensional counterpart \cite{kl}. In this
framework, the vortex/soliton states are related to the associated free
fermions. This modified Maxwell theory also appears as the effective
three dimensional theory describing the real electromagnetic interaction
experienced by four dimensional charged particles constrained to move on
an infinite plane \cite{qedpl}. 

In order to obtain massive vortex states, we consider an operator which 
is dual to the vortex operator \cite{lcs} and construct a gas containing
the point charges created by it in the grand-canonical ensemble. A new
theory is thereby generated, whose vacuum functional is the grand-partition
function of the gas. In this theory, which we call ``Sine-Maxwell'', in
analogy to the Sine-Gordon, the  charge bearing dual operator has a nonzero
vacuum expectation value, resembling what happens when the gauge theory
is coupled to a symmetry breaking Higgs field. Of course, one immediately  
would ask whether a mass would be dynamically generated to the gauge field
in this theory. In Section 4 we explain that further investigation is 
required in order to answer this question. 

We evaluate the vortex two-point function in the above described theory
using a fugacity expansion for the gas. Comparing the resulting series
with the corresponding one for the Sine-Gordon solitons we can immediately
infer the massiveness of the quantum vortices.

\section{ Vortex Operators}
\setcounter{equation}{0}

Quantum vortex creation operators in 2+1D were introduced years ago in
theories containing an abelian vector gauge field \cite{pol,vor,nv}.
In a pure Maxwell theory, namely,
\be
\cl_M = -\fr{1}{4} F^{\mu\nu} F_{\mu\nu}
\label{m}
\ee
it is given by \cite{nv,ap}
\be
\mu(x) \exp \lef \{ -ib \int_{T_{L(x)}} d^2 \xi  
\del_i arg(\xi -x) F^{i0}(\xi,x^0) \ri \} 
\label{mu}
\ee
or 
\be
\mu(x;T_{L(x)}) = \exp\lef \{ -{i \over 2}\int d^3z  A^{\mu\nu}
F_{\mu\nu}\right \}
\label{vorqed2}
\ee
where the external field $A^{\mu\nu}$ is given by 
\be
A_{\mu\nu} (z;x) = b \int_{ T_{L(x)}} d^2\xi_{[\mu}\del_{\nu]}
 \arg(\vec \xi -\vec x)\delta^3 (z-\xi)
\label{Amn}
\ee
Where $T_{L(x)}$ is the surface represented in Fig.1. Even though $\mu$
depends explicitly on the surface $T_{L(x)}$, it can be shown \cite{vor,nv}
that its renormalized correlation functions are local. The vortex operator
$\mu$ creates quantum states carrying $2\pi b$ units of flux.
In this trivial case the renormalized quantum vortex correlation                                 
functions are \cite{nv,ap}
\be
<\mu \mu^\dagger>_R = \exp \lef\{ \fr{\pi b^2}{|x-y|} 
 \ri \}
\label{cf6}
\ee
The large distance behavior indicates that $<\mu>_R \neq 0$ and therefore
the $\mu$-operator does not create genuine vortex excitations in pure
Maxwell theory as one would expect.

A nonlocal generalization of Maxwell theory given by
\be
\cl_{MNL} = -\fr{1}{4} F^{\mu\nu}\lef[\fr{1}{(-\Box)^{1/2}}\ri ] F_{\mu\nu}
= -\fr{1}{4}\int d^3y F^{\mu\nu}(x)K(x-y) F_{\mu\nu}(y)
\label{mnl}
\ee
has been studied recently in connection with the bosonization of a free 
massless Dirac fermion field in 2+1D \cite{bos}. This theory was also shown 
to describe the actual electrodynamic interaction of charged particles 
constrained to move on an infinite plane \cite{qedpl}. In (\ref{mnl})
\be
K(x - y) =   
\lef[\fr{1}{(-\Box)^{1/2}}\ri ] =
\int {d^3 k\over{(2\pi)^3}} {e^{ik\cdot(x - y)}\over
{(k^2 + i\epsilon)^{1/2}}}
= {i\over{2\pi\left[(x - y)^2  
 + i \epsilon \right]}}
\label{kernel}
\ee
The quantization of theories of
the type given by (\ref{mnl}) has been studied carefully in \cite{ru}.
In spite of the nonlocality, they are shown to respect causality and to
be perfectly well defined. In the specific case of (\ref{mnl}), the 
Green functions are shown to have support on the light-cone surface
\cite{ru}, a fact which makes the 
theory to obey the Huygens principle in analogy to the
four dimensional Maxwell theory, a property which is not shared by 
its three dimensional counterpart.
 
In this case, the vortex operator $\mu$ is given by
\cite{bos,lcs}
\be
\mu(x;T_{L(x)}) = \exp\lef \{ -{i \over 2}\int d^3z  A^{\mu\nu}
\lef[\fr{1}{(-\Box)^{1/2}}\ri ]F_{\mu\nu}\right \}
\label{vorqed3}
\ee
where $A^{\mu\nu}$ is still given by (\ref{Amn}).
The renormalized correlation function of $\mu$ in the theory described by
(\ref{mnl}) has been evaluated in \cite{lcs}. The result is
\be
<\mu(x) \mu^\dagger(y)>_R =  \fr{1}{|x-y|^{b^2}} 
\label{cf66}
\ee
The large distance behavior of (\ref{cf66}) now indicates that 
$<\mu> =0$ and therefore there are true quantum vortex excitations in
this theory. The power-law decay, on the other hand, is an evidence that
these states are massless. These massless vortex states are the ones
that are associated to the massless fermion field in the process of 
bosonization introduced in \cite{bos}. The situation is analogous to the
one occuring in 1+1D when the soliton operator introduced in the theory
of a massless scalar field creates massless states which correspond to
the massless fermion field \cite{cm}. The theory described by (\ref{mnl})
is the three dimensional analog of the 1+1D massless scalar field.
In order to obtain a theory with nontrivial massive vortex excitations
what is done in the two dimensional case is that an order operator dual to
$\mu$ is introduced \cite{odd} and a theory is constructed whose vacuum
functional is the grand partition function of a gas of the excitations
created by this order parameter \cite{sgcg}. The resulting theory is the 
familiar Sine-Gordon \cite{cm}. In the following Section, we are
going to perform the analogous steps in 2+1D. We will start from (\ref{mnl})
and consider the order operators dual to $\mu$ which were introduced in 
\cite{bos,lcs}. Then, we will construct a gas with the excitations created
by these operators and eventually arrive at a partition function that
is associated to a theory which is the three dimensional counterpart of
the Sine-Gordon.

\section{ The ``Sine-Maxwell'' Theory}
\setcounter{equation}{0}

\subsection{The $\sigma$ Operator and Its Correlation Functions}

The $\sigma$ operator which is dual to $\mu$ in 
the nonlocal Maxwell theory described by the 
lagrangian (\ref{mnl}) was introduced and 
studied in \cite{bos,lcs}. Similarly to $\mu$, we
can also write it in terms of an external field, which in this case is 
given by 
\be
\C_{\mu\nu}(z;x) =  a \int_{x,L}^\infty d\xi_\mu 
\del_\nu \lef[\fr{1}{(-\Box)^{1/2}}\ri] (z-\xi)
\label{c1}
\ee
We can express $\sigma$ as a functional of $\C_{\mu\nu}$ as \cite{lcs}
$$
\sigma(x) = \exp \lef \{- i \int d^3z \C_{\mu\nu}
\lef[\fr{1}{(-\Box)^{1/2}}\ri]F^{\mu\nu}
 \ri \}
$$
or
$$
\sigma(x) = \exp \lef \{ ia \int_{x,L}^\infty d\xi_\beta \int d^3z
\lef [\fr{1}{-\Box}\ri](\xi -z) \del_\alpha F^{\alpha\beta}(z)
 \ri \} 
$$
\be
\sigma(x) = \exp \lef \{ ia \int_{x,L}^\infty d\xi^i A_i(\xi) 
+ 
\lef [\fr{\del_\alpha A^\alpha (x)}{-\Box}\ri] 
 \ri \} 
\label{si}
\ee
We see that $\sigma$ is gauge invariant. As in the case of $\mu$
the renormalized correlation functions of $\sigma$ are local,
in spite of the fact that it explicitly depends on the curve $L$ \cite{lcs}.
It can be shown 
\cite{lcs} that $\sigma$
creates states bearing $a$ units of the charge corresponding to the 
conserved current 
\be
j^\mu = \del_\nu \lef [\fr{F^{\nu\mu}}{(-\Box)^{1/2}}\ri ]
\label{j}
\ee
The renormalized euclidean correlation functions of $\sigma$ are given
by \cite{lcs}
$$
<\sigma(x)\sigma^\dagger(y)> =
Z^{-1} \int DA_\mu \expo 
\lef[ \fr{1}{4}
F_{\mu\nu}\lef[\fr{1}{(-\Box)^{1/2}}\ri ]
F^{\mu\nu}
\ri.\ri.
$$
\be
\lef.\lef.
+i\  F_{\mu\nu} \lef[\fr{1}{(-\Box)^{1/2}}\ri ] 
\C^{\mu\nu}
- \fr{1}{2} \C_{\mu\nu}\lef[\fr{1}{(-\Box)^{1/2}}\ri ] 
\C^{\mu\nu}  \ri]\ri\}
\label{cf77}
\ee
where $\C^{\mu\nu}=\C^{\mu\nu}(z;x)-\C^{\mu\nu}(z;y)$. An arbitrary
n-point correlation function would be obtained by the introduction
of additional external sources $\C^{\mu\nu}(z;x_i)$.
The second term in the above expression corresponds to the operator itself
and the last one is a renormalization factor which makes the 
correlation function completely local (independent of the path L).
Expression (\ref{cf77}) was evaluated in \cite{lcs} giving the result
\be
<\sigma(x)\sigma^\dagger(y)> = \lim_{m,\epsilon\rightarrow 0}
\exp\left\{ - \fr{a^2}{4\pi^2}\left[\ln m 
\vert x - y \vert - \ln m \vert \epsilon \vert \right] \right\}
\label{ssnr}
\ee
where $m$ and $\epsilon$ are respectively infrared and ultraviolet
regulators introduced in order to control the singularities
of the euclidean propagator corresponding to (\ref{mnl}),
namely
$$
D^{\sigma\lambda} = \lef (-\Box \delta^{\sigma\lambda}+ 
\lef ( 1 - \fr{1}{\xi} \ri )\del^\sigma\del^\lambda
\ri ) \lef [\fr{1}{(-\Box)^{3/2}}\ri] 
$$
with
\be
\left[{1\over{(-\Box)^{3/2}}}\right](x - y) =
\lim_{m, \epsilon \rightarrow 0} - {1\over{8\pi^2}}\ln m^2\left[ 
\vert x - y \vert^2 + \vert \epsilon \vert^2\right]
\label{prop}
\ee
($\xi$ is the gauge fixing parameter). 

We see that the infrared cutoff $m$ is completely cancelled and we get
\be
<\sigma(x)\sigma^\dagger(y)>_R = {1\over{\vert x - y \vert^\fr{a^2}{4\pi^2}}}
\label{ssr}
\ee
where the ultraviolet cutoff was eliminated through the renormalization
$\sigma_R = \sigma e^{\vert \epsilon^2 \vert /2}$. Should we compute 
a non-neutral correlation function like $<\sigma\sigma>$ we would have
the relative sign in (\ref{ssnr}) reversed, the $m$ factors would no longer
be cancelled and the correlation function would be equal to zero.

Another interesting result is the mixed $\sigma-\mu$ correlation function.
This was also evaluated in \cite{lcs}, giving the result
$$
<\sigma(x_1)\sigma^\dagger(x_2)\mu(y_1)\mu^\dagger(y_2)>_R =  
\fr{1}{ |x_1-x_2|^{\fr{a^2}{4\pi^2}}}  \fr{1}{ |y_1-y_2|^{b^2}}   
\exp \lef\{               
A(x_1 -y_1) + A(x_1 -y_2) + \ri.
$$
\be
\lef.
A(x_2 -y_1) + A(x_2 -y_2) \ri\}
\label{cf11}
\ee
where
\be
A(x_i -y_j) \equiv n\ iab \  arg(x_i -y_j)
\label{ct}
\ee
with $n=0,\pm 1,\ldots$. The ambiguity of the above correlation function
up to $arg(x_i -y_j)$ functions is a reflex of the many possible orderings
of operators in the l.h.s. and indicates that $\sigma$ and $\mu$ do
satisfy the following dual algebra \cite{lcs}
\be
\sigma(x)\mu(y) = \mu(y)\sigma(x)\exp 
\left\{i a b\, \arg(\vec y - \vec x )
\right\}
\label{sm}
\ee

\subsection{The ``Sine-Maxwell'' Theory}

In order to obtain a pure gauge theory with massive vortex excitations 
in 2+1D, let us proceed as in the case of Sine-Gordon and
consider the theory whose vacuum functional is given by
\be
Z_R = \sum_{m = 0}^\infty (-1)^m {(\alpha/2)^m\over{m!}}  
\int \prod_{i = 1}^m d^3z_i 
\sum_{\lambda_i} 
Z_R^{(m)}(z_1, \ldots z_{m}) 
\label{z1}
\ee
where 
\be
Z_R^{(m)}(z_1, \ldots z_{m})=
<\sigma_{\lambda_1}(z_1) \ldots \sigma_{\lambda_m}(z_m)>_{\alpha=0,R} 
\label{z11}
\ee
and $\alpha$ is a real parameter which will be later on identified
with the fugacity of the gas.
In (\ref{z1}) and (\ref{z11}) $\lambda_i = \pm 1$ and 
we use the convention that $\sigma_+ =\sigma$
and $\sigma_- =\sigma^\dagger$. The sum $\sum_{\lambda_i}$ runs over  
all possible configurations of $\lambda$'s in the set $\{\lambda_i\}$.
As we saw in the last subsection, only ``neutral'' configurations 
for which $\sum_i^m \lambda_i =0 $ and
$m=2n$, do contribute to (\ref{z1}). 
The $2n$-point correlation
function
$$
<\sigma_{\lambda_1}(z_1) \ldots \sigma_{\lambda_n}(z_{2n})>_{\alpha=0,R} =
\lim_{m,\epsilon \rightarrow 0} 
\exp\left\{ {a^2\over {16\pi^2}}\sum_{i,j=1}^2n 
\lambda_i \lambda_j \ln m^2 \left[\vert z_i - z_j \vert^2 
+\vert \epsilon \vert^2
\right] \right\} 
$$
\be
= Z_R^{(2n)}(z_1, \ldots z_{2n})
\label{z2}
\ee
is the Boltzmann weight of a three dimensional gas of point particles
with a logarithmic interaction. From (\ref{z2}) it is clear that the 
infrared cutoff $m$ is completely canceled because 
$\sum_i^m \lambda_i =0 $.
Since $m=2n$, the sum $\sum_{\lambda_i}$ in (\ref{z1}) 
just gives an overall factor 
$\fr{m!}{(n!)^2}$. The self-energies of the particles, 
which correspond to the 
$i=j$ terms in (\ref{z2}) can be eliminated by renormalizing the 
fugacity as $\alpha_R = \alpha |\epsilon|^{\fr{a^2}{8\pi^2}}$. The 
resulting vacuum functional is the grand-partition function of the same 
gas of point particles with logarithmic interaction, namely,
\be
Z_R =
 \sum_{n=0}^\infty {(\alpha_R/2)^{2n}\over{(n!)^2}} 
\int \prod_{i=1}^{2n} d^3 z_i 
\exp\left\{{a^2\over {8}\pi^2} {\sum_{i\neq j=1}^{2n}} 
\lambda_i \lambda_j \ln [\vert z_i -
z_j \vert  + \vert \epsilon \vert] \right\}
\label{z3}
\ee
This is almost identical to the two-dimensional Coulomb Gas of point
particles \cite{sgcg,lm}. Here, however, the logarithmic interaction
is not a Coulomb interaction.
There are still ultraviolet divergences in 
(\ref{z3}) caused by the coalescence of $p$-positive and $p$-negative
charges. The analysis of 
these divergences can be made exactly 
by the same method applied to the Coulomb Gas
in the 1+1D case \cite{lm}. We will have divergences for
\be
a^2 =8\pi^2 {3 (2 p - 1)\over{2p}}    
\label{div}
\ee
corresponding to the coalescence of a neutral $p$-pole.
For $a^2 < 12\pi^2$, the theory is finite because the 
singularities are integrable. 
For $12\pi^2 \leq  a^2 < 24\pi^2$ we will have the divergences
corresponding to neutral $p$-poles, given by (\ref{div}). 
These can be eliminated 
by a multiplicative renormalization of (\ref{z3}) \cite{lm}.   
For $a^2 \geq 24\pi^2$, we start to have divergences corresponding
to non-neutral configurations  
in addition to the ones given by (\ref{div}) and we do not know how 
to eliminate them \cite{lm}. We can assert, however, that the theory
makes sense for $a^2 < 24\pi^2$. At this point the theory probably 
undergoes a phase transition analogous to the one of Kosterlitz-Thouless 
which takes place in the corresponding two-dimensional system \cite{lm}.

Let us investigate now whether we can associate a lagrangian to the 
vacuum functional $Z_R$. For this let us consider the 
functional integral expressing the $m$-point $\sigma$ correlation
function
$$
Z_R^{(m)}(z_1, \ldots z_{m})   = 
Z_0^{-1} \int DA_\mu \exp\left\{ - \int d^3z 
\left[{1\over 4}F^{\mu\nu}\lef[\fr{1}{(-\Box)^{1/2}}\ri ] F_{\mu\nu}  
+ \cl_{GF}  
\right. \right.   
$$
$$
  + i\  \sum_{i=1}^m 
\lambda_i \C_{\mu\nu}(z;z_i)\lef[\fr{1}{(-\Box)^{1/2}}\ri ] F_{\mu\nu}
$$
\be
 -
\left. \left.\left[ \sum_{i = 1}^m \lambda_i \tilde C_{\mu\nu}(z;z_i)\right]
\left[{1\over{(-\Box)^{1/2}}}\right]
\left[ \sum_{j = 1}^m \lambda_j\tilde C_{\mu\nu}(z;z_j)\right]\right]\right\}
\label{z4}
\ee
where $\C_{\mu\nu}$ is given by (\ref{c1}).
Neglecting the last term which is a renormalization factor, and inserting 
the unrenormalized Boltzmann weight $Z^{(m)}(z_1, \ldots z_{m})$ in 
(\ref{z1}) we get
$$
Z = \sum_{m = 0}^\infty (-1)^m {(\alpha/2)^m\over{m!}}  
\int \prod_{i = 1}^m d^3z_i 
\sum_{\lambda_i} 
Z_0^{-1} \int DA_\mu \exp\left\{ - \int d^3z 
\left[{1\over 4}F^{\mu\nu}\lef[\fr{1}{(-\Box)^{1/2}}\ri ] F_{\mu\nu}  
\right. \right.  
$$
$$
\left. \left. + \cl_{GF} + i\  \sum_{i=1}^m 
\lambda_i \C_{\mu\nu}(z;z_i)\lef[\fr{1}{(-\Box)^{1/2}}\ri ] F_{\mu\nu}
\right]\right\} 
$$
We immediately recognize the sum in (\ref{z1}) as an
expression of the cossine and therefore 
$$
Z =
Z_0^{-1} \int DA_\mu \exp\left\{ - \int d^3z 
\left[{1\over 4}F^{\mu\nu}(z)\lef[\fr{1}{(-\Box)^{1/2}}\ri ] F_{\mu\nu}(z)  
+ \cl_{GF}  
\right. \right. + 
$$ 
\be
 \lef.\lef.
\alpha \cos \lef [ \int d^3z' \C_{\mu\nu}(z;z')
\lef[\fr{1}{(-\Box)^{1/2}}\ri ] F^{\mu\nu}(z') \ri ] \ri ] \ri \}
\label{z5}
\ee
The theory which corresponds to
$Z$ is therefore described by the lagrangian
\be
\cl_{SM} =
{1\over 4}F^{\mu\nu}\lef[\fr{1}{(-\Box)^{1/2}}\ri ] F_{\mu\nu} 
+\alpha \cos \lef [ \int d^3z \C_{\mu\nu}
\lef[\fr{1}{(-\Box)^{1/2}}\ri ] F^{\mu\nu} \ri ] 
\label{sm}
\ee
This is the three-dimensional theory which corresponds to the Sine-Gordon.
Since it involves the vector gauge field we call it ``Sine-Maxwell''. In
its renormalized form, given by definition by the 
functional $Z_R$, Eq. (\ref{z3}), it 
becomes equivalent to a three-dimensional gas of point particles with
a logarithmic interaction. As we mentioned before this is perfectly
well defined for a certain range of the parameter $a$, namely
$a^2 < 24\pi^2$.

In the next section, we are going to show that this theory presents
some features that are usually associated to the presence of a Higgs
field with a symmetry breaking potential coupled to a gauge field, namely,
a non-vanishing vacuum expectation value for the charge bearing operator,
in this case $\sigma$, and the presence of massive excitations carrying
the topological charge, that is, massive vortices.

\section{ Correlation Functions}
\setcounter{equation}{0}   

\subsection{$<\sigma>$}

Let us consider firstly the vacuum expectation value of the charge 
bearing operator $\sigma$. As we saw in Section 2, 
we had $<\sigma> = 0$ 
in the nonlocal Maxwell theory, given by (\ref{mnl})
because of the infrared divergences appearing
in (\ref{ssnr}) and (\ref{prop}). In the ``Sine-Maxwell theory, we
have
\be
<\sigma(x)> = 
\sum_{m = 0}^\infty (-1)^m {(\alpha/2)^m\over{m!}}  
\int \prod_{i = 1}^m d^3z_i 
\sum_{\lambda_i} 
Z_R^{(m)}(x;z_1, \ldots z_{m}) 
\label{cf1}
\ee
where $Z_R^{(m)}(x;z_1, \ldots z_{m}) = 
<\sigma(x)\sigma_{\lambda_1}(z_1) \ldots 
\sigma_{\lambda_m}(z_m)>_{\alpha=0,R}  $ 
is given by an expression identical to
(\ref{z4}), except for the exchange
\be
\sum_{i = 1}^m \lambda_i \tilde C_{\mu\nu}(z;z_i) \longrightarrow
\tilde C_{\mu\nu}(z;x) + \sum_{i = 1}^m \lambda_i \tilde C_{\mu\nu}(z;z_i)
\label{ex}
\ee
As only neutral configurations can contribute to $<\ldots>^{\alpha=0}$ we
immediately see that now we must have $m=2n+1$ with $n$ positive 
and $n+1$ negative $\lambda$'s. The expression for 
the appropriate free correlation function appearing 
in (\ref{cf1}) is a trivial extension of (\ref{z2}) 
\cite{lcs}:
$$
Z_R^{(2n+1)}(x;z_1, \ldots z_{n};\overline z_1 \ldots \overline z_{n+1})=
\lim_{m,\epsilon \rightarrow 0}
\exp\left\{\fr{a^2}{16\pi^ 2}\left[
2 \sum_{i=1}^n \ln m^2 [\vert x  - z_i \vert^2 + \epsilon^2] 
\ri.\ri.
$$
$$
\lef.\lef.
- 2 \sum_{i=1}^{n+1} \ln m^2
[\vert x - \overline z_i \vert ^2 + \epsilon^2]
+ \ln m^2 \epsilon^2 + \ri.\ri. 
$$
\be
\lef.
+ \sum_{i,j=1}^n 
\ln m^2 [\vert z_i - z_j \vert^2 + \epsilon^2] 
+ \sum_{i,j=1}^{n+1} \ln m^2 [\vert \overline z_i - 
\overline z_j \vert^2 + \epsilon^2] + 
- 2 \sum_{i=1}^n \sum_{j=1}^{n+1} \ln m^2
[\vert z_i - \overline z_j \vert^2 + \epsilon^2] \right\}
\label{cf2} 
\ee
We immediately see that the infrared cutoff $m$ is proportional to
$(2n-2(n+1)+1+n^2+(n+1)^2-2n(n+1))=0$ and therefore completely disappears.
Renormalizing $\alpha$ as before, introducing the renormalized $\alpha$:
$\alpha_R = \alpha \epsilon^{a^2/8\pi^2}$ and inserting the resulting
correlation function
in (\ref{cf1}), we get
$$
<\sigma(x)>_R = \sum_{n=0}^\infty {(\alpha_R/2)^{2n + 1}\over{(n!)(n + 1)!}} 
\int \prod_{i=1}^{2n} d^3 z_i 
\int \prod_{i=1}^{2n+1} d^3 \overline z_i 
\exp\left\{{a^2\over8\pi^ 2}\left[
2 \sum_{i=1}^n \ln  \vert x  - z_i \vert 
\ri.\ri.
$$
\be
\lef.\lef.
- 2 \sum_{i=1}^{n+1} \ln 
\vert x - \overline z_i \vert 
+ \sum_{i\neq j=1}^n 
\ln \vert z_i - z_j \vert + \sum_{i\neq j=1}^{n+1} \ln \vert \overline z_i - 
\overline z_j \vert + 
- 2 \sum_{i=1}^n \sum_{j=1}^{n+1} \ln  
\vert z_i - \overline z_j \vert \ri] \right\}
\label{cf3} 
\ee
Since the infrared cutoff $m$  has been completely canceled 
from the above expression we conclude that
$<\sigma(x)>_R \neq 0$. This happens because the extra charge introduced
by the operator $\sigma$ has been neutralized by the ``gas charges'' in
the grand-canonical ensemble. The situation is analogous to the one
which occurs with a Higgs field in the presence of a symmetry breaking
potential. It would be very interesting to investigate whether this
fact would produce a massive state in the gauge field spectrum as in the 
Higgs mechanism. However, the situation here is not as clear because
the mass spectrum of nonlocal theories like (\ref{sm}) is quite unusual
\cite{ru}.

\subsection{$<\mu(x) \mu^\dagger(y)>$}

Let us turn now to the $<\mu\mu^\dagger>$ correlation function. It
is clear that
\be
<\mu(x) \mu^\dagger(y)>_R= 
\sum_{m = 0}^\infty (-1)^m {(\alpha/2)^m\over{m!}}  
\int \prod_{i = 1}^m d^3z_i 
\sum_{\lambda_i} 
<\mu(x) \mu^\dagger(y)\sigma_{\lambda_1}(z_1) \ldots 
\sigma_{\lambda_m}(z_m)>_{\alpha=0,R} 
\label{cf4}
\ee
The above mixed correlation functions were calculated in \cite{lcs}.
A special case of them is given by (\ref{cf11}). The neutrality condition
is now clearly respected with an equal number of positive and negative gas
charges. The fugacity $\alpha$ is renormalized as before. Inserting the
general expression for the mixed correlation functions in (\ref{cf4})
\cite{lcs} (also according to (\ref{cf11})) we obtain
$$
<\mu(x)\mu^\dagger(y)>_R =  \sum_{n=0}^\infty {(\alpha_R/2)^{2n}
\over{(n!)^2}} 
\int \prod_{i=1}^{2n} d^3 z_i 
\exp \left\{ - b^2  \ln \vert x - y \vert  
+{a^2\over2} \sum_{i\neq j}^{2n} \overline \la_i \overline \la_j \ln \vert 
z_i - z_j \vert 
\ri.
$$ 
\be
\lef.
+{iab \over4} \sum_{i=1}^2 \sum_{j=1}^{2n} \la_i \overline \la_j 
A(z_j -  x_i) \right\}
\label{cf5}
\ee
where $A(z_j -  x_i)$ was defined in (\ref{ct}). From the above expression,
we see that 
\be
\lim_{\vert \vec x - \vec y \vert \rightarrow \infty} <\mu(x)\mu^\dagger(y)> 
= 0
\label{cf6}
\ee
indicating that $<\mu>=0$ which implies by its turn that the $\mu$-operator
does create true excitations out of the vacuum. Comparing (\ref{cf5})
with the corresponding expression in the Sine-Gordon theory \cite{sg1}
we can see that they are identical, except for the dimensionality of the
space. On the basis of this observation we should expect that the vortex
states created by $\mu$ are massive as the corresponding soliton 
excitations in the Sine-Gordon theory.

\section{Conclusion}
\setcounter{equation}{0}   

Starting from a gauge theory in which the vortex operator creates massless
states in 2+1D, we made a parallel with the theory of a free massless scalar 
field in two dimensions, where the quantum solitons have zero mass, and 
used a gas of point charges construction, in order to obtain a pure abelian
gauge theory which in many senses resembles the Sine-Gordon. The quantum 
vortices become massive in this theory in analogy to the Sine-Gordon 
solitons. Also there is a symmetry breaking order parameter analog to the 
vacuum expectation value of a Higgs or Sine-Gordon field. The theory is
completely mapped in a three dimensional gas of point charges with 
logarithmic interaction in the grand-canonical ensemble. The analysis of
divergences which was applied to the Coulomb gas in 2D can be carried
through straightforwardly in the 3D case and we thereby can conclude that 
the theory is sensible for a certain range of the coupling constant.

The starting 
theory which one obtains by turning off the cossine interaction appears
in the bosonization of the free massless Dirac fermion field in 2+1D. We
therefore should expect that the theory obtained when the cossine interaction 
is included may be related to the bosonization of the massive fermion field 
in three dimensional space-time. This point deserves a further investigation.
Also an interesting subject wich is presently being investigated is the 
existence of classical vortex solutions with a finite energy. These would
be the classical counterparts of the massive quantum states created by the 
vortex operator.

\bigskip                
\leftline{\Large\bf Acknowledgements} \bigskip

Both authors  
were partially supported by CNPq-Brazilian National Research Council.
FIT is very grateful to Department of Physics and Astronomy of 
the University of Pittsburgh for the kind hospitality.

\end{document}